\input{aipcheck}

\documentclass[
    ,final            % use final for the camera ready runs
  ]
  {aipproc}

\layoutstyle{6x9}

\begin{document}

\title{The overview of the spin physics at RHIC-PHENIX experiment}

\classification{13.85.Ni, 13.88.+e, 21.10.Hw, 25.40.Fq, 14.20.Dh, 13.60.Hb}
\keywords      {proton, spin structure, asymmetry}

%<Replace this text with PACS numbers; choose from this list:
%                \texttt{http://www.aip..org/pacs/index.html}>

\author{Yoshinori Fukao for the PHENIX Collaboration}{
address={Department of Physics, Kyoto University, Kyoto, 606-8502, Japan}
}

%\author{}{
%  address={<common address for author2 and author3>}
%}

%\author{<author3>}{
%  address={<common address for author2 and author3>}
%  ,altaddress={<author1 address>} % additional visiting address
%}

\begin{abstract}

In Spiring 2005, RHIC successfully completed its first long data collection
run with polarized proton beams. PHENIX accumulated ten fold larger
statistics with higher polarization than the previous spin physics run
in 2003. This contribution will introduce the RHIC-PHENIX experiment
and present our recent results.

\end{abstract}

\maketitle

%%%%%%%%%%%%%%%%%%%%%%%%%%%%%%%%%%%%%%%%%%%%
%% MAINMATTER
%%%%%%%%%%%%%%%%%%%%%%%%%%%%%%%%%%%%%%%%%%%%

\vspace{-1.0cm}
\section{Introduction}
\vspace{-0.3cm}
One of the major goals of the PHENIX spin program is to experimentally
study the origin of the proton spin. Past measurements in polarized
deep inelastic scattering (DIS) have led to the so-called ``spin crisis'':
quarks and anti-quarks carry only $\sim30$\% of the nucleon
spin\cite{r:AAC_paper}. Contributions from gluon spin and orbital angular
momentum of partons remain unknown.

In the past we reported the double helicity asymmetry ($A_{LL}$) for
inclusive $\pi^0$ production\cite{r:run3_pi0_ALL}. The measurement of
the $A_{LL}$ is one of the promising approaches to the gluon polarization
in the proton ($\Delta g$).
In the 2005 run, RHIC has operated for 3 months for spin physics
with center-of-mass energy ($\sqrt{s}$) of 200 GeV and average
beam polarization of 47\%.
PHENIX accumulated an integrated luminosity
of $\sim$ 3.8 pb$^{-1}$ with longitudinal polarization and $\sim$ 0.1 pb$^{-1}$
with transverse polarization. The greatly improved RHIC performance
allows a wide-range of measurements using different probes including pions,
electrons, muons, prompt photons and jets as well as the $\pi^0$s.

In addition to the measurement of $\Delta g$, studies in progress
include single transverse-spin asymmetries ($A_N$), spin transfer to
the $\Lambda$ and J/$\psi$ polarization.
Towards the highest energy run at $\sqrt{s}=500$ GeV planned in near future,
polarized beams were successfully accelerated and brought into collisions
at $\sqrt{s}=410$ GeV in the 2005 run.
%For the future plan, accelerating
%and colliding beams at $\sqrt{s}=410$ GeV was also tested in the 2005 run.

\vspace{-0.4cm}
\section{PHENIX detector}
\vspace{-0.3cm}
The PHENIX detector\cite{r:phenix_detector} consists of global detectors and
four spectrometers.
One of the global detectors is a set of beam-beam counters (BBC) for the
determination of the vertex position of the collision. The minimum bias
trigger (MB)
is provided by BBC and the number of MB events is used as the relative
luminosity for the asymmetry analysis.
The other detector is a zero degree calorimeter
(ZDC) located at very forward region. The ZDC also plays an important role
in the polarized proton run as the local polarimeter, which monitors
the direction of the beam polarization by measuring the neutron $A_N$
and the longitudinal component was more than 98\% during the 2005 run.

Two spectrometers out of four are constructed in the central region
(central arm) covering $-0.35$ - $+0.35$ in pseudorapidity
and azimuthal angle of $180^{\circ}$ in total. The central arm
consists of tracking detectors, particle identification detectors and
electromagnetic calorimeters (EMCal) at the end. Charged particles and photons
can be detected by these arms. The high energy photon trigger by means of
the EMCal is implemented and it is utilized to collect high transverse
momentum ($p_T$) photons, $\pi^0$s and jets.
The other two spectrometers are composed of three stations of
tracking chambers and a 5-layer sandwich of chambers and a steel to
identify muons (muon arm).

\vspace{-0.3cm}
\section{Recent results}
\vspace{-0.3cm}
In this conference, many talks from PHENIX are presented.
The latest results from the 2005 run include single transverse-spin
asymmetry of charged pions\cite{r:run5_pion_AN} and double helicity
asymmetry of $\pi^0$\cite{r:run5_pi0_ALL} and J/$\psi$\cite{r:run5_Jpsi_ALL}.
The study of helicity correlated difference in jet intrinsic $k_T$
to study the orbital angular momentum of the parton is
presented\cite{r:run3_jet_kT_asym}.
Many other works are ongoing and some of them are introduced in this report.

\begin{figure}[htbp]
\begin{minipage}{0.6\linewidth}
\begin{center}
\includegraphics[width=0.98\linewidth]{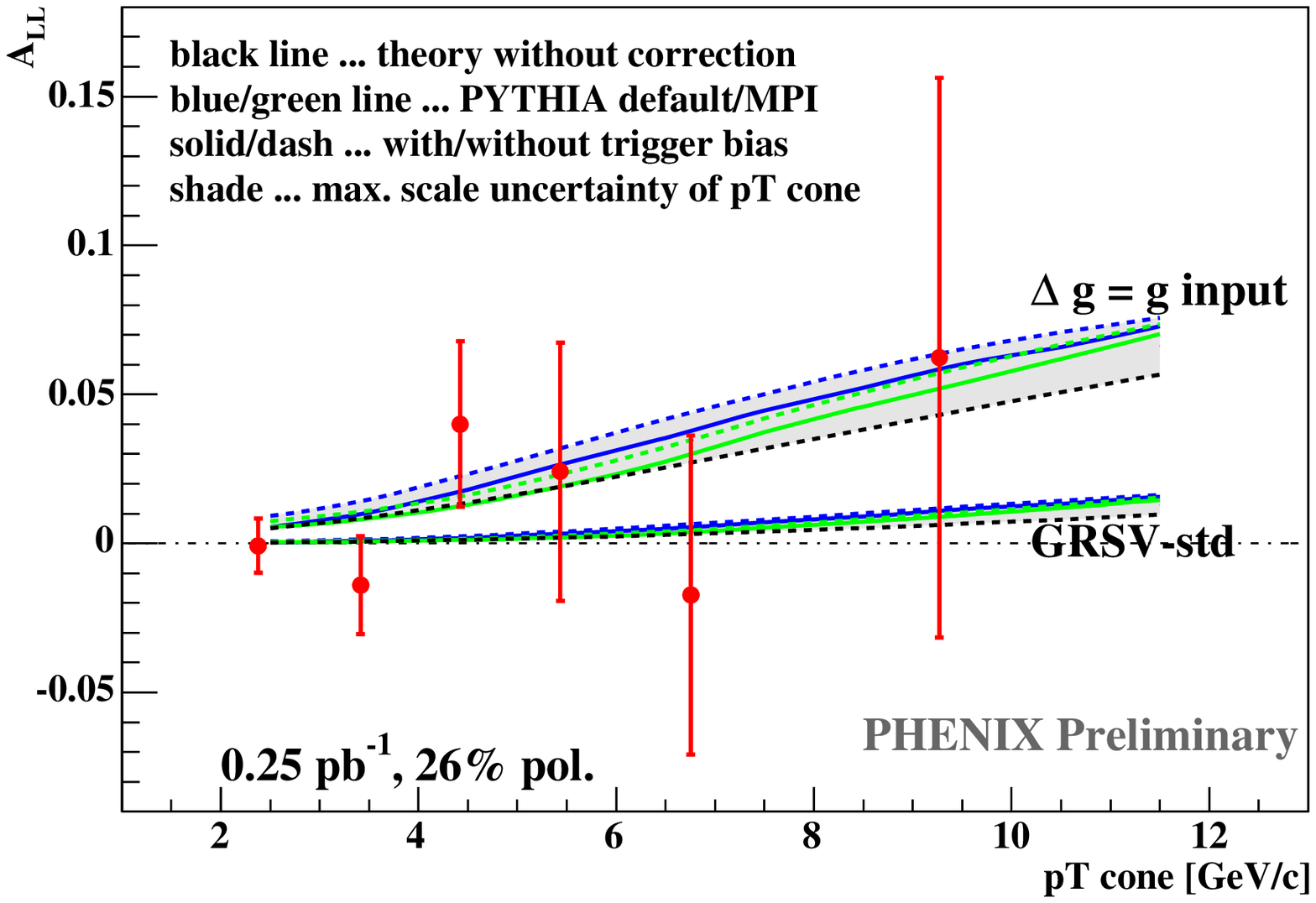}
\end{center}
\end{minipage}
\begin{minipage}{0.4\linewidth}
\begin{center}
\includegraphics[width=0.9\linewidth]{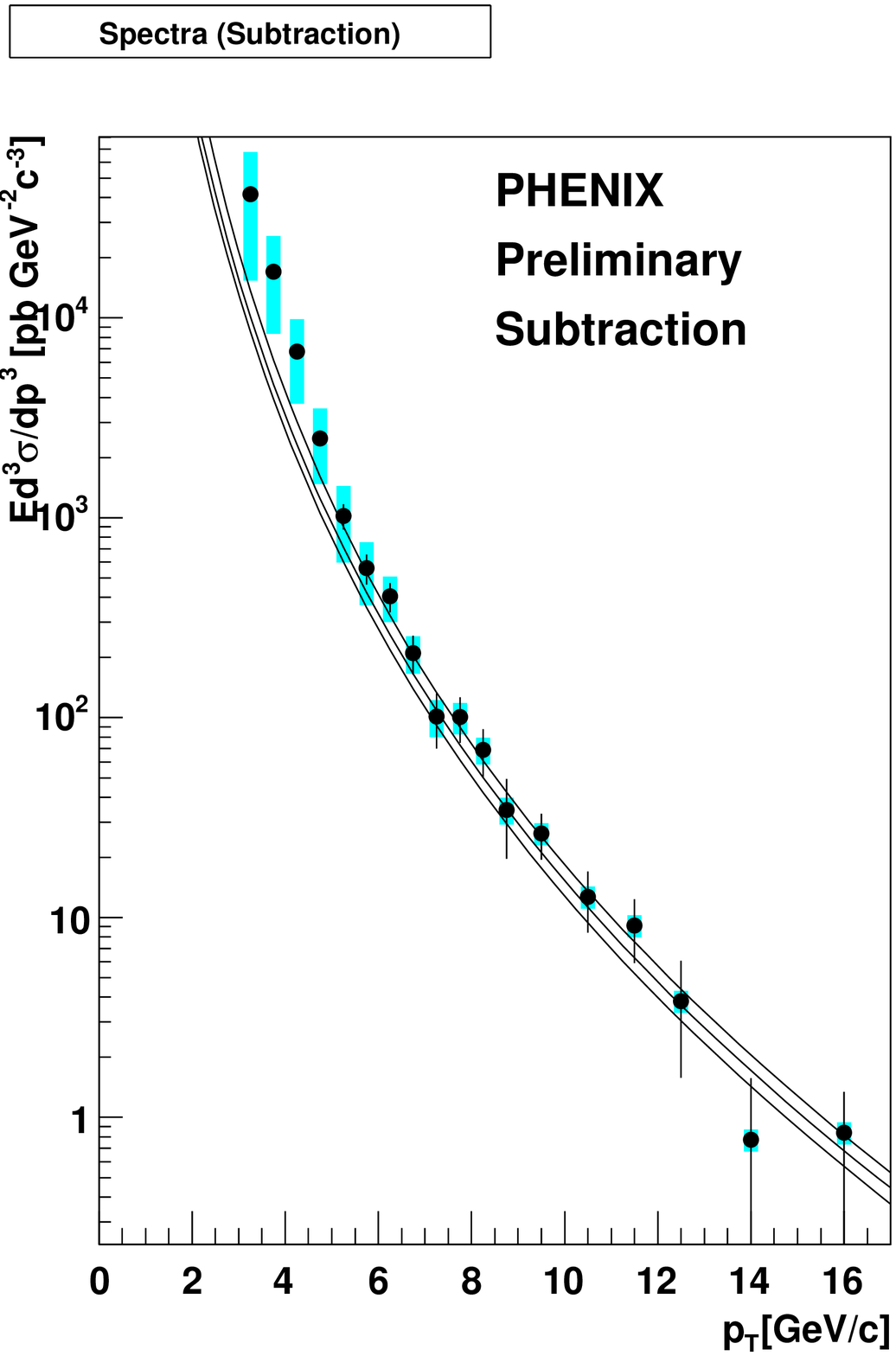}
\end{center}
\vspace{-0.5cm}
\end{minipage}
\caption{(a)-left; $A_{LL}$ of the jet as a function of the observed $p_T$
($p_T$ cone). The theoretical curves are normalized based on the PYTHIA
simulation. Black; without scaling (original), Blue; scaled using 
default PYTHIA,
Green; scaled using PYTHIA with the correction of multi-parton
interaction which reproduces the data better than default PYTHIA.
The bias of high energy photon trigger is (not) considered
in solid (dashed) line. (b)-right; The cross section of prompt photons.
The black bars and blue bands
denote the statistical and systematic uncertainty respectively. The line
is the NLO pQCD calculation with scale of $p_T / 2$ (upper curve), $p_T$
and $2 p_T$ (lower curve).}
\label{p:jet_and_dp}
\end{figure}

One of the important channels is the prompt photon to study $\Delta g$.
While the $\pi^0$ $A_{LL}$ with high statistics strongly constrains the size of
$\Delta g$, it is difficult to determine the sign of the
$\Delta g$. The $\pi^0$ $A_{LL}$ can be roughly described as the
quadratic equation of $\Delta g$ since $\pi^0$ production with high $p_T$
originates in $gg$, $gq$ and $qq$ scatterings, where $g$ ($q$) denotes
gluon (quark). This ambiguity can be resolved by
measuring $A_{LL}$ for the prompt photon production which is dominated
by $gq$ scattering.
Figure \ref{p:jet_and_dp} (b) shows the cross section of the prompt
photon measured by the central arms in the 2003 run with the theoretical
calculation based on next leading order (NLO) perturbative quantum
chromodynamics (pQDC). The theory describes the data well at the $p_T>5$ GeV.
Because of the low statistics, it is necessary to accumulate more data with
high luminosity and polarization to obtain precision results of the asymmetry.

Another effective probe to determine
$\Delta g$ is the $A_{LL}$ measurement of inclusive jet production.
In addition to high statistics, the advantage over the $\pi^0$ is that
jet is free from the uncertainty of the fragmentation. In the analysis,
the jet is measured by collecting the photons and charged particles
within the cone of 0.3
% $\sqrt{|\phi|^2 + |\eta|^2} = 0.3$ 
in the central region.
The fraction of the jet momentum observed by the detector to the
initial momentum of the scattered parton is evaluated by the simulation
to be $\sim 80$\%.
Figure \ref{p:jet_and_dp} (a) shows the results of the jet $A_{LL}$ in the
2003 run. Two theoretical curves are also shown, one using the best
global fit to the inclusive DIS data (GRSV-std) and another one using
a polarized gluon distribution equal to the unpolarized distribution
($\Delta g = g$)\cite{r:jet_ALL_theory}. The horizontal axis is
the jet $p_T$ observed by the PHENIX detector ($p_T$ cone). The theoretical
curves are scaled along horizontal axis based on the PYTHIA simulation.
The precision of the data will be much improved by the data taken in
the 2005 run.

\vspace{-0.3cm}
\section{For the future}
\vspace{-0.3cm}
For the future program, it is planned to extract polarized
anti-quark distribution through the $W$ bosons, whose coupling is flavor
sensitive and quarks and anti-quarks can be separated. It is necessary
to operate RHIC with higher $\sqrt{s}$
%(The goal is $\sqrt{s}=500$ GeV)
to accumulate enough statistics of $W$. In the 2005 run, RHIC tested
the operation at $\sqrt{s}=410$ GeV and we confirmed that the beam
is accelerated with significant polarization by measuring
single transverse-spin asymmetry for the forward neutron production.

New hardware is being developed for the future plan.
One of them is the silicon vertex tracker to make the acceptance of jet
wider and tag the heavy flavor particle by detecting the displaced vertex.
The trigger upgrade for the muon arms is in progress for the $W$ detection by
tagging high momentum muons.

\vspace{-0.3cm}
\section{Summary}
\vspace{-0.3cm}
PHENIX collected the data in the first long polarized proton run in
2005. The measurement of the double helicity asymmetry in $\pi^0$ production
constrains $\Delta g$. It is expected that
the study of other channel will help the systematic understanding about
$\Delta g$. Many other efforts, including the measurement of $A_N$, jet $k_T$,
spin transfer of $\Lambda$ and J/$\psi$ polarization, are underway.
The commissioning of the polarized proton-proton collision at
$\sqrt{s}=410$ GeV was successfully done in the 2005 run.
As well as the accelerator development, the detector upgrades in
the PHENIX is ongoing for the future programs.

%\bibliographystyle{aipproc}   % if natbib is available
%\bibliographystyle{aipprocl} % if natbib is missing

%%%%%%%%%%%%%%%%%%%%%%%%%%%%%%%%%%%%%%%%%%%
%% You probably want to use your own bibtex database here
%%%%%%%%%%%%%%%%%%%%%%%%%%%%%%%%%%%%%%%%%%%
%\bibliography{sample}

%%%%%%%%%%%%%%%%%%%%%%%%%%%%%%%%%%%%%%%%%%%
%% Just a reminder that you may have to run bibtex
%% All of it up to \end{document} can be removed
%% if you don't like the warning.
%%%%%%%%%%%%%%%%%%%%%%%%%%%%%%%%%%%%%%%%%%%
%\IfFileExists{\jobname.bbl}{}
% {\typeout{}
%  \typeout{******************************************}
%  \typeout{** Please run "bibtex \jobname" to optain}
%  \typeout{** the bibliography and then re-run LaTeX}
%  \typeout{** twice to fix the references!}
%  \typeout{******************************************}
%  \typeout{}
% }

\end{document}